\documentstyle[12pt,aps,epsf]{revtex}

\newcommand{\bce}{\begin{center}}
\newcommand{\ece}{\end{center}}
\newcommand{\beq}{\begin{equation}}
\newcommand{\eeq}{\end{equation}}
\newcommand{\be}{\begin{equation}}
\newcommand{\ee}{\end{equation}}
\newcommand{\bea}{\begin{eqnarray}}
\newcommand{\eea}{\end{eqnarray}}

\newcommand{\ba}{\begin{array}}
\newcommand{\ea}{\end{array}}

\newcommand{\doublespace}{
    \renewcommand{\baselinestretch}{1.6}\large\normalsize}

\def\lsim{\mathrel{\rlap{\lower4pt\hbox{\hskip1pt$\sim$}}
    \raise1pt\hbox{$<$}}}         
\def\gsim{\mathrel{\rlap{\lower4pt\hbox{\hskip1pt$\sim$}}
    \raise1pt\hbox{$>$}}}         

\def\Pom{{\bf I\!P}}

 \advance\oddsidemargin   by -1in
 \advance\evensidemargin  by -1in
 \advance\topmargin   by -0.5in
\begin{document}



\doublespace
\title{
Predictions for the forward cone in diffractive DIS
\vspace{2.0cm}\\}

\author{N.N.~Nikolaev$^{a,b}$,
A.V.Pronyaev$^{c}$,
B.G.~Zakharov$^{b}$}
\address
{$^{a}$IKP(Theorie), FZ J{\"u}lich, D-52425 J{\"u}lich, Germany
\medskip\\
$^{b}$L. D. Landau Institute for Theoretical Physics, GSP-1,
117940, \\
ul. Kosygina 2, Moscow V-334, Russia.
\medskip\\
$^{c}$Department of Physics,
Virginia Polytechnic Institute and State University,\\
Blacksburg, VA 24061, USA\vspace{2.0cm}\\}
\maketitle

\begin{abstract}
We calculate the diffraction slope $B_{D}$ for diffractive DIS.
We find a
counterintuitive rise of $B_{D}$ from exclusive
diffractive excitation of vector mesons to excitation of continuum
states with $M^{2} \sim Q^{2}$. For the small-mass continuum we predict 
a rapid variation of $B_{D}$ with $M^{2}$ on the scale $m_{V}^{2}$
and a sharp drop of $B_{D}$ for a small-mass continuum above the 
vector meson excitation. 

\end{abstract}


\newpage

\doublespace

The diffraction slope is one of the principal observables which measures
the impact parameter structure of diffractive scattering.  The
comissioning of the leading proton spectrometer (LPS) of the ZEUS detector 
at HERA \cite{ZEUSLPS} gave a long awaited access to the transverse 
momentum transfer $\vec{\Delta}$ and the diffraction slope $B_{D}=
-\partial \log\left\{d\sigma_{D}/d\Delta^{2})\right\}/\partial \Delta^{2}$ 
in diffractive deep inelastic scattering (DIS) $ep\rightarrow e'p'X$. 
The special interest in diffraction slope for diffractive DIS stems from 
the fact that besides the mass $M$ of the excited state $X$ there emerges 
a new large scale: the virtual photon's mass  $\sqrt{Q^{2}}$. The principal 
issue is what $B_{D}$ depends from: $M^{2}, Q^{2}$, the mass $m_{V}$ 
of the ground-state vector meson in the corresponding flavour channel 
and/or the 
diffractive scaling variable $\beta=Q^{2}/(Q^{2}+M^{2})$ (hereafter 
$Q^{2}, x $ and $x_{\Pom}=x/\beta$ are the standard diffractive DIS 
variables).

This is a highly nontrivial issue because at fixed $\beta$ diffraction
proceeds into the high-mass continuum states $X$ with $M^{2} = Q^{2}(1-\beta)
/\beta \gg m_{V}^{2}$. Our experience with diffraction of hadrons and/or real
photons can be summarized as follows. For any two-body diffractive scattering 
$ac\rightarrow bd$, an essentially model-independent decomposition holds, 
$B_{D}=\Delta B_{ab}+\Delta B_{cd} + \Delta B_{int}, $ 
where $\Delta B_{ij}$ comes from the size of the $ij$ transition vertex 
and the relatively small $\Delta B_{int}$ comes from the interaction 
range proper \cite{HoltmannDD,DDhadronic}. The values of $\Delta B_{ij}$ 
depend strongly on the excitation energy in the $i\rightarrow j$ transition, 
$\Delta M^{2} =m_{j}^{2}-m_{i}^{2}$. In elastic scattering, $i=j$, one 
finds $\Delta B_{ii}\approx {1\over 3}R_{i}^{2} \sim$ 4-6 GeV$^{-2}$, 
where $R_{i}^{2}$ is the mean squared hadronic radius, and typically 
$B_{el}\sim 10$ GeV$^{-2}$. The similar estimate $\Delta B_{ij}\approx 
{1\over 3}R_{i}^{2}, {1\over 3}R_{j}^{2}$ holds for diffraction into 
low-mass  continuum states, $\Delta M^{2} \lsim m_{N}^{2}$, and 
diffraction into low-mass continuum and elastic scattering fall into 
the broad category of {\sl exclusive} diffraction for which $B_{D} \sim 
B_{el}$. However, for excitation of high-mass continuum, $\Delta M^{2} 
\gsim m_{N}^{2}$, often referred to as the triple-pomeron ($3\Pom$) and/or
genuine {\sl inclusive} region of diffraction, the size of the diffracting
particle no longer contributes to the diffractive slope 
and $B_{D}=B_{3\Pom} =\Delta B_{pp} +\Delta B_{int} 
\sim {1\over 2}B_{el} \approx 6$ GeV$^{-2}$. The above slope $B_{3\Pom}$ 
is about universal for all the diffracting beams and excited states 
$X$ \cite{DDhadronic}). 
Furthermore, in the double high-mass diffraction $hp \rightarrow XY$,
when $M_{X,Y} \gg m_{N}$, one is left with very small 
$B_{D} \sim \Delta B_{int}
\sim $1.5-2 GeV$^{-2}$ (\cite{HoltmannDD,Conta} and references therein).
In real photoproduction the excitation scale is definitely set
by the ground-state vector meson mass $m_{V}$.  
Perhaps the most dramatic example of this distinction between 
{\sl exclusive} and {\sl inclusive} diffraction is a drastic change of the 
diffraction slope from elastic, $pA\rightarrow pA$, to quasielastic,
$pA\rightarrow p'A^{*}$, scattering of protons on heavy nuclei 
\cite{Glauber}. 

Another well understood diffractive process is elastic production of
vector mesons $\gamma^{*}p\rightarrow p'V$. In this case the transverse
size $\gamma^{*}\rightarrow V$ transition vertex, the so-called 
scanning radius 
\be
r_{S}= {6 \over \sqrt{Q^{2}+m_{V}^{2}}}
\label{eq:Rscan}
\ee
decreases with $Q^{2}$ (and $m_{V}^{2}$). This is a basis of the prediction
\cite{NNPZZ98} of $\Delta B_{\gamma^{*}V}\propto r_{S}^{2}$ 
and of the decrease of the diffraction slope $B_{V}$ down to $B_V\approx 
B_{3\Pom}$ at very large $Q^{2}$, which is in good agreement with 
the experiment \cite{ZEUSVM98}. 

In this paper we report predictions for  
the $Q^{2},M^{2}$ and flavour dependence of the diffraction 
slope for inclusive diffractive DIS. We demonstrate that in striking 
contrast to $B_{V}$ for exclusive diffraction into vector 
mesons which exhibits strong
dependence on $Q^{2}$, the diffraction slope $B_{D}$ for inclusive
diffractive DIS is a scaling function of $\beta$. The most paradoxical
prediction is that in contrast to real photon and hadronic diffraction,
in diffractive DIS $B_{D}$ rises with the excited mass $M$ reaching
$B_{D} \sim B_{el}$ at $M^{2}\sim Q^{2}$. Arguably, such an unusual 
behaviour of $B_{D}$ derives from the scaling scanning radius $r_{S}$ for
diffraction excitation of continuum  $q\bar{q}$  states 
\cite{GNZcharm},
\be
r_{S}^{2}  \sim {9\over m_{f}^{2}} (1-\beta)\, ,
\label{eq:RsDDIS}
\ee
which rises towards small $\beta$, so that $\Delta B_{\gamma^{*}X} 
\propto r_{S}^{2}$ does not depend on $Q^{2}$ and rises substantially 
from $\beta \approx 1$ to $\beta \sim {1\over 2}$.  Earlier such a 
large, $Q^{2}$-independent $\Delta B_{\gamma^{*}X}$ has been conjectured 
in \cite{NZ94} and in the present communication we quantify this
property of the diffraction slope by a direct calculation. Furthermore, 
we predict a substantial drop of $B_{D}$ below $B_{3\Pom}$ for excitation 
of the small-mass continuum.

Finally, 
for very large excited masses, $M^{2} \gg Q^{2}$, i.e., $\beta \ll 1$, 
even for the $q\bar{q}$ excitation one recovers the inclusive regime 
of small $\Delta B_{\gamma^{*}X}$ and $B_{D}$ decreases back to 
$B_{D}\sim B_{3\Pom}$. This triple-pomeron limit of $\beta \ll 1$
is dominated by excitation of the $q\bar{q}g$ and higher Fock states
of the photon, though, which is the genuinely {\sl inclusive} process 
and by the same token as for hadronic diffraction one
can argue \cite{NZ94} that $B_{D}$ must not depend on $Q^{2}$ and
that $B_{D} \approx B_{3\Pom}$. This anticipation has been confirmed by
the first data from the ZEUS LPS: $B_{D}= 7.2\pm 1.1(stat.)^{+07}_{-0.9}
(syst.)$ GeV$^{-2}$ for diffractive DIS ( $5<Q^{2}< $20 GeV$^{-2}$)
\cite{ZEUSLPS} and $B_{D}=6.8 \pm 0.9(stat.)^{+1.2}_{-1.1}(syst.)
$GeV$^{-2}$ in real photoproduction ($Q^{2}=0$) \cite{ZEUSdifreal}.

We focus on diffractive excitation of the $q\bar{q}$ Fock states of the 
photon, which is known to dominate at $\beta \gsim 0.1$ \cite{GNZ95}.
The sample Feynman diagram for this process is shown in Fig.1, in which we 
show also all the relevant momenta. We base our analysis on the formalism 
\cite{NZsplit}, which we generalize to the non-forward
case $\vec{\Delta}\neq 0$. \footnote{The first 
calculation of the diffraction slope for the $M^{2}$ integrated cross 
section is found in \cite{Diploma}, the preliminary results from the 
present study have been reported elsewhere \cite{DIS98}}. 

If $z$ and $(1-z)$ are fractions of the (lightcone) momentum of the photon
carried by the quark and antiquark, respectively and $\vec{k}$ is the
relative transverse momentum in the $q\bar{q}$ pair, then
$M^2={m_{f}^2 + k^2 \over z(1-z)}$. The quark and antiquark are produced 
with the transverse momenta $\vec{k}+z\vec{\Delta}$ and
$-\vec{k}+(1-z)\vec{\Delta}$ with respect to the $\gamma^{*}p$ collision 
axis. We focus on the transverse diffractive structure function (SF).
To the leading log${1\over x_{\Pom}}$, for
excitation of quarks of mass $m_{f}$ and electric charge $e_{f}$,
\bea
F_{T}^{D(4)}(\vec{\Delta}\,^{2},x_{\Pom},\beta,Q^{2})=
{8\pi e_{f}^{2} \over 3\sigma_{tot}(pp)}
\int {d^{2}\vec{k} \over 2\pi} {(k^{2}+m_{f}^{2})\beta \over
(1-\beta)^{2}J}
\alpha_{S}^{2}(\overline{Q}^{2})
\left\{\left[1-2z(1-z)\right] \vec{\Phi}_{1}^{2}  +
m_{f}^{2}\Phi_{2}^{2}
\right\} \, ,
\label{eq:FT}
\eea
where $J=\sqrt{1-4(k^{2}+m_{f}^{2})/M^{2}}$, $\alpha_{S}^{2}(\overline{Q}^{2})$
is the strong coupling, evaluated at the QCD hardness
scale $\overline{Q}^{2}$ to be
specified below, and $f\left(x_{\Pom},\vec{\kappa},\vec{\Delta}\right)$
is the gluon density matrix \cite{NNPZZ98,Lipatov}. In the 
calculation of diffractive helicity 
amplitudes $\vec{\Phi}_{1},\Phi_{2}$ it is convenient to introduce
\be
\psi(z,\vec{k})= {1\over \vec{k}^{2}+m_{q}^{2}+z(1-z)Q^{2}}\,, ~~
\vec{\Psi}(z,k)=\vec{k}\psi(z,\vec{k})\,,
\ee
in terms of which
\bea
\Phi_i=\int {d^2\vec{\kappa} \over 2\pi\kappa^{4}}
f\left(x_{\Pom},\vec{\kappa},\vec{\Delta}_{\perp}\right)
\phi_{i}
\label{eq:PHII}
\eea
where
\bea
\vec{\phi}_{1}=\vec{\Psi}(z,\vec{r}+\vec{\kappa})
+\vec{\Psi}(z,\vec{r}-\vec{\kappa})
-\vec{\Psi}(z,\vec{r}+{1\over 2}\vec{\Delta})-
\vec{\Psi}(z,\vec{r}-{1\over 2}\vec{\Delta})\, ,
\label{eq:phi1}\\
 \phi_{2}=\psi(z,\vec{r}+\vec{\kappa})
+\psi(z,\vec{r}-\vec{\kappa})
-\psi(z,\vec{r}+{1\over 2}\vec{\Delta})-
\psi(z,\vec{r}-{1\over 2}\vec{\Delta})\, ,
\label{eq:phi2}\\
\vec{r}=\vec{k}-{1\over 2}\left(1-2z\right)\vec{\Delta}
\label{eq:r}
\eea
For small $\vec{\Delta}$ within the diffraction cone 
\begin{equation}
{\cal{F}}(x,\vec{\kappa},\vec{\Delta})=
{\partial G(x,\kappa^{2})\over \partial \log \kappa^{2}}
\exp(-{1\over 2}
B_{3\Pom}\vec{\Delta}^{2})\,. 
\label{eq:24}
\end{equation}
where  $\partial G/\partial \log \kappa^{2}$ is the conventional 
unintegrated gluon structure function  \cite{NNPZZ98}. The dependence of 
${\cal{F}}(x,\vec{\kappa},\vec{\Delta})$ on $\vec{\Delta}\vec{\kappa}$ 
corresponds to the subleading BFKL singularities \cite{Lipatov} and 
can be neglected at small $x_{\Pom}$. The diffraction slope $B_{3\Pom}$
in (\ref{eq:24}) is a nonperturbative quantity, it comes for the most 
part form the hadronic size of the proton, modulo to a slow Regge 
growth one can take $B_{3\Pom}\sim$ 6 GeV$^{-2}$ \cite{NNPZZ98}.

In the present analysis we are mostly concerned with the $\beta,Q^{2}$
and flavour dependence of $\Delta B_{\gamma^{*}X}$ which comes from the
$\vec{\Delta}$ dependence of $\vec{\phi}_{1}$ and $\phi_{2}$, for
our purposes it is sufficient to evaluate $\vec{\Phi}_{1}^{2},\Phi_{2}^{2}$
to an accuracy $\vec{\Delta}^{2}$. The calculation of amplitudes 
$\vec{\Phi}_{1},\Phi_{2}$ has been discussed to great detail in 
\cite{GNZcharm,NZsplit,BGNPZ98} and need not be repeated
here. We simply cite the results starting with excitation of heavy
quark-antiquark pair, when the fully  perturbative QCD (pQCD)
analytic calculation is possible:
\bea
F_{T}^{D(4)}(t,x_{\Pom},\beta,Q^{2})=\frac{2\pi
e_f^2}{9\sigma_{tot}(pp)}
\frac{\beta\left(1-\beta\right)^2}{m_f^2}
\left[
\left(3+4\beta+8\beta^2\right)
\right.
\nonumber \\
\left.
+\frac{\Delta^2}{m_f^2}\cdot\frac{1}{10}
\left(5-16\beta-7\beta^2-78\beta^3+126\beta^4\right)\right]\cdot
\left[\alpha_{s}(\overline{Q}^{2})G(x_{\Pom},\overline{Q}^{2})\right]^{2}
\exp\left(-B_{\Pom}\vec{\Delta}^{2}\right)\, .
\label{eq:S3T}
\eea
where the pQCD hardness scale equals
\be
\overline{Q}^{2} \approx m_{f}^{2}(1+{Q^{2} \over M^{2}})=
{m_{f}^{2} \over 1-\beta}\, .
\label{pQCDscale}
\ee
The result (\ref{eq:S3T}) holds for the large-mass continuum, 
$M^{2} \gg 4m_{f}^{2}$. 
As it has been shown in \cite{GNZcharm}, the typical transverse size 
in the $\gamma^{*}\rightarrow q\bar{q}$ transition vertex is 
$1/\overline{Q}$ , see eq. (\ref{eq:RsDDIS}). For excitation of heavy
flavors and/or for light flavours at $1-\beta \ll 1$ the hardness
scale $\overline{Q}^{2}$ is large and one is in the legitimate pQCD 
domain. 

Consequently, the contribution to the diffraction slope from the
$\gamma^{*}X$ excitation vertex equals
\be
\Delta B_{\gamma^{*}X}=
{1\over m_f^2}\cdot {16\beta+7\beta^2+78\beta^3-126\beta^4-5 \over
10(3+4\beta+8\beta^2)}
\label{eq:DeltaB}
\ee
which is a rigorous pQCD result for heavy flavours.
Evidently, it is a scaling function of $\beta$ which does not depend 
on $Q^{2}$, which nicely correlates with the scanning radius being
a function of $\beta$ only. It rises from $\beta\sim 1$ to $\beta 
\sim {1\over 2}$ and decreases in the inclusive limit of $\beta \to 0$.
It diminishes the diffraction slope at $\beta \sim 1$, which can
be attributed to the $s$-channel helicity nonconserving spin-flip 
transitions.

One can readily evaluate $\Delta B_{\gamma^{*}X}$ for the both terms
$\propto \vec{\Phi}_{1}^{2}$ and $m_{f}^{2}\Phi_{2}^{2}$, we only 
comment here that for the both terms the $\beta$ dependence of 
 $\Delta B_{\gamma^{*}X}$ is very similar to that
given by eq. (\ref{eq:DeltaB}).
Even for heavy flavours, the contribution to  $F_{T}^{D(4)}$ from 
$m_{f}^{2}\Phi_{2}^{2}$ is a numerically small correction to the dominant
contribution from $\propto \vec{\Phi}_{1}^{2}$. This correction
is even smaller for lighter flavours. As it has been discussed 
in \cite{BGNPZ98}, the scale $\mu_{G}$ of variation of the unintegrated 
gluon density in the soft-to-hard transition region becomes more important
than the mass $m_{f}$ of light quarks. For this reason, for light flavour
excitation the contribution from $m_{f}^{2}\Phi_{2}^{2}$ will be suppressed
$\propto m_{f}^{2}/\mu_{G}^{2}$. Furthermore, the scale for 
$\Delta B_{\gamma^{*}X}$ will be set by $1/\mu_{G}^{2}$ rather than
by $1/m_{f}^{2}$. One of the consequences is that the change of 
$\Delta B_{\gamma^{*}X}$ from strange to up/down quarks is much
weaker than $\propto 1/m_{f}^{2}$, see fig.~2 where we show our
numerical results.

Although for light flavours the magnitude of $\Delta B_{\gamma^{*}X}$ 
is no longer pQCD calculable,
the behaviour of the unintegrated gluon density in the soft-to-hard 
transition region is reasonably well tested from earlier calculations
\cite{BGNPZ98}
of the diffractive SF $F_{T}^{D(4)}$ which agree with the experiment, 
and also from the small-$Q^{2}$ behaviour of the proton structure function
\cite{NZZF2}. The emergence of this second scale has only a 
marginal impact on the $\beta$-dependence of $B_{D}$ what we here are 
concerned about. We checked that variations of $B_{D}$ calculated 
using different soft-to-hard interpolations of the gluon structure 
function as described in \cite{BGNPZ98} do not exceed $\sim 1$ GeV$^{-2}$ 
with the retention of the form of the $\beta$ dependence of $B_{D}$.

In contrast to the scaling $\beta$ dependence of $B_{D}$ for finite 
$\beta$, for diffractive DIS into near-threshold small masses, 
$M^{2} \sim m_{V}^{2}\sim 4m_{f}^{2}$, i.e., for $1-\beta \propto 
{M^{2}\over Q^{2}} \ll 1$, we predict a strong $M^{2}$ dependence 
of the diffraction slope. The near-threshold region belongs to the 
pQCD domain even for light flavour excitation, because here the 
QCD hardness scale is large, 
$\overline{Q}^{2} \approx {1\over 4}(Q^{2}+m_{V}^{2})$.( For finite 
$Q^{2}$ and/or heavy flavours one must bear in mind the kinematical 
threshold $\beta \leq \beta_{th}=Q^{2}/(Q^{2}+4m_f^2)< 1$.) The plane 
wave description of final states holds for the quark-antiquark relative 
velocity $v\gsim \alpha_{S}(\overline{Q}^{2})$.  
In this case the small-$v^{2}$ expansion of diffractive SF is
\be
F_{T}^{D(4)}(t,x_{\Pom},v,Q^{2})=\frac{128\pi e_f^2}{3\sigma_{tot}(pp)}
{m_f^2 \over Q^4} v \left[1+
{\Delta^2 \over 6m_f^2}v^2\right]\cdot 
\left[\alpha_{s}(\overline{Q}^{2})G(x_{\Pom},\overline{Q}^{2})\right]^{2}
\exp\left(-B_{\Pom}\vec{\Delta}^{2}\right)\, .
\label{eq:VT}
\ee
The principal effect is that the diffraction slope {\sl decreases}
with the increase of $v^{2}$ and/or $M^{2}$:
\be
\Delta B_{\gamma^{*}X} = -{v^{2}\over 6m_{f}^{2}}\, .
\label{eq:DeltaB2}
\ee
Here for heavy flavours $v^{2} = 1 -{4m_{f}^{2} \over M^{2}}$, for
light flavours it only makes sense to speak of the continuum above
the ground-state $1S$ vector mesons ($\rho^{0},\omega,\phi^{0}$) and
$v^{2}$ must be understood as $v^{2} \sim 1 - {m_{V}^{2} \over M^{2}}$.
Consequently, for the small-mass continuum we predict very rapid 
variations of the diffraction slope $B_{D}$, see fig.~2, and here 
the relevant mass scale is $m_{V}^{2}$. The principal point is that 
$B_{D}$ drops substantially, we leave open the scenario in which $B_{D}$ 
becomes negative valued, i.e., there will be a forward dip, in a certain
range of masses. 
   
In the spirit of duality for diffractive DIS \cite{GNZlong}, diffraction 
excitation of the small-mass continuum above the $1S$ ground state 
vector meson is dual to production of radial excitations of vector 
mesons. Then, our finding of the near-threshold decrease of the 
diffraction slope with rising $M^{2}$ correlates nicely with the 
prediction that for the $V'(2S)$ states the diffraction slope is 
substantially smaller than for the ground state vector mesons $V(1S)$, 
which follows from the node effect \cite{NNPZZ98} . The 
near-threshold drop of $B_{D}$ is smaller for heavy flavours, in a nice
conformity with the weaker node effect in diffractive production of
heavy quarkonia.

The similar analysis can be repeated for the longitudinal diffractive
structure function. Although it is of higher twist, it dominates 
diffractive DIS at $\beta \gsim 0.9$ \cite{GNZlong,BGNPZ98}. As far as
the diffraction slope is concerned, the QCD hardness scale for
diffraction excitation of longitudinal photons is large,
$\overline{Q}^{2} \approx {1\over 4\beta}Q^{2}$, the corresponding 
scanning radius is small and we expect $B_{D} \approx B_{3\Pom}$.

We conclude with a somewhat academic observation on a sum rule for 
the $M^{2}$ integrated cross section of diffractive excitation of 
heavy $q\bar{q}$ pairs by transverse photons. Namely, if one neglects 
the $\beta$ dependence of the QCD hardness scale $\overline{Q}^{2}$ 
in (\ref{eq:S3T}), then one readily finds that for the $M^{2}$-integrated 
diffractive cross section $\Delta B_{\gamma^{*}X} = 0$ and $B_{D}=B_{3\Pom}$. 
Indeed, a closer inspection of the calculation of the $M^{2}$ integrated 
cross section shows that to the accuracy $\Delta^{2}$ the dependence 
on $\vec{\Delta}$ can be eliminated by the change of the integration 
variable $d^2\vec{k} \rightarrow d^2\vec{r}$. One can trace the origin 
of this sum rule to a QCD gauge invariance properties of (\ref{eq:PHII}), 
it serves as a useful cross check of corresponding polynomial coefficients. 
This sum rule is of little practical value, though, because for the
dominant excitation of light flavours $\overline{Q}^{2}$ is small, in the 
soft-to-hard transition region of a strong variation of the gluon 
structure function $G(x_{\Pom},\overline{Q}^{2})$ and the above
outlined derivation is not applicable.

To summarize, we presented predictions from the standard two-gluon 
pomeron exchange
mechanism for the forward cone in diffractive DIS. For the high-mass
continuum excitation we predict that the diffractive slope $B_{D}$ is 
a scaling function of $\beta$ which has a counterintuitive rise from 
small masses to $M^{2} \sim Q^{2}$, which has no analogue in diffraction
of real photons and/or hadrons. For the small-mass continuum we predict 
a rapid variation of $B_{D}$ with $M^{2}$ on the scale $m_{V}^{2}$
and a 
sharp drop of $B_{D}$ for a small-mass continuum above the vector meson
excitation. These predictions can be tested at HERA.

The work of B.G.Z. has been supported partly by the INTAS grant 96-0597
and the work of A.V.P. was supported partly by the US DOE grant 
DE-FG02-96ER40994.

{\bf Figure caption:}\\

Fig.1: One of the four Feynman diagrams for diffraction excitation
of the $q\bar{q}$ final state via QCD two-gluon pomeron 
exchange.\\

Fig.2: Our predictions for the $\beta$ and flavour dependence of the
diffraction slope $B_D$ in diffractive DIS of transverse photons
at $Q^{2}=100$ GeV$^{2}$.


\begin{thebibliography}{299}


\bibitem{ZEUSLPS} 
ZEUS Collab. J. Breitweg, M.Derrick, D.Krakauer et al. {\sl Eur. Phys. J.}
{\bf C1}, 109 (1998)

\bibitem{HoltmannDD} 
Holtmann H., N.N.Nikolaev, J.Speth, A.Szczurek and B.G.Zakharov, 
{\sl Z. Phys. } {\bf C69}, 297 (1996).

\bibitem{DDhadronic} 
Alberi G. and G.Goggi, {\sl Phys. Rep.} {\bf 74}, 1 (1981); 
Chapin T.J., R.L. Cool, K. Goulianos et al., {\sl Phys. Rev. }
{\bf D31}, 17 (1985).

\bibitem{Conta} 
Conta C., M. Fraternali, F. Gigli-Berzolari et al.,
{\sl Nucl. Phys. } {\bf B175}, 97 (1980).

\bibitem{Glauber} 
R.J. Glauber and G. Matthiae, {\sl Nucl. Phys.} {\bf B21}, 135 (1970)

\bibitem{NNPZZ98} 
Nemchik J.,N.N.Nikolaev, E.Predazzi, B.G.Zakharov and V.R.Zoller,
{\sl J.Exp.Theor.Phys.} {\bf  86} (1998) 1054;
Nemchik J., N.N. Nikolaev, E. Predazzi and B.G. Zakharov, {\sl Z. Phys.}
{\bf C75}, 71 (1997); Nikolaev N.N., B.G.Zakharov and V.R.Zoller,
{\sl Phys. Lett.} {\bf B366} (1996) 337



\bibitem{ZEUSVM98} 
ZEUS Collab. J.Breitweg, S.Chekanov, M.Derrick et al., DESY-98-107, 
hep-ex/9808020.


\bibitem{GNZcharm} 
Genovese M., N.Nikolaev  and B.Zakharov,
{\sl Phys. Lett.} {\bf B378}, 347 (1996).

\bibitem{NZ94} 
Nikolaev N.N. and B.G.Zakharov,  {\sl J. Exp. Th. Phys.}
{\bf 78}, 598 (1994);
{\sl Z. Phys.} {\bf C64} (1994) 631.


\bibitem{ZEUSdifreal} 
ZEUS Collab. J. Breitweg, M.Derrick, D.Krakauer et al.  {\sl Eur. Phys. J.}
{\bf C2}, 237 (1998)

\bibitem{GNZ95} 
Genovese M., N.N. Nikolaev and  B.G Zakharov, {\sl J. Exp. Theor. Phys.}
{\bf 81}, 625 (1995). 

\bibitem{NZsplit} 
Nikolaev N.N. and B.G.Zakharov, {\sl Phys. Lett.} {\bf B332}, 177 (1994); 
{\it Z. Phys.} {\bf C53},  331 (1992).

\bibitem{Diploma} 
Pronyaev A.V. and B.G.Zakharov (1992), unpublished. A.V.Pronyaev,
Diploma Thesis, Moscow Institute for Physical Engineering (1992)


\bibitem{DIS98} 
N.N.Nikolaev and B.G.Zakharov, Phenomenology of Diffractive DIS,
in: Deep Inelastic Scattering and QCD (DIS97), Proc. of 5th Intenational
Workshop, Chicago, IL USA, April 14-18, 1997, American Institute
of Physics  Proceedings No.407, edited by J.Repond and D.Krakauer,
pp. 445-455; A.Pronyaev, The Forward Cone and L/T Separation in
Diffractive DIS, hep-ph/9808432, To be published in the proceedings 
of 6th International Workshop on Deep Inelastic Scattering and QCD (DIS 98),
Brussels, Belgium, April 4-8, 1998. 


\bibitem{Lipatov} 
L.N.Lipatov, 
{\sl Sov. Phys. JETP} {\bf 63} (1986) 904; 
L.N.Lipatov, in:{\sl Perturbative Quantum Chromodynamics}, 
ed. by A.H.Mueller, World Scientific (1989); 
E.A.Kuraev, L.N.Lipatov and S.V.Fadin, 
{\sl Sov. Phys. JETP} {\bf 44} (1976) 443; 
{\sl Sov. Phys. JETP} {\bf 45} (1977) 199;  


\bibitem{BGNPZ98} 
Bertini M., M. Genovese, N.N. Nikolaev, A.V. Pronyaev and B.G. Zakharov, 
{\sl Phys. Lett.} {\bf B422}, 238 (1998).


\bibitem{NZZF2} 
N.N. Nikolaev, B.G. Zakharov and V.R. Zoller, 
{\sl JETP Lett.} {\bf 66}, 138 (1997), {\sl Pisma Zh. Eksp. Teor. Fiz.}
{\bf 66},  134 (1997); N.N.Nikolaev and B.G.Zakharov, 
{\sl Phys.Lett.} {\bf B327}, 157 (1994).

\bibitem{GNZlong} 
Genovese M., N.Nikolaev and B.Zakharov, {\sl Phys.Lett.}
{\bf B380},  213 (1996).


 



\end{thebibliography}
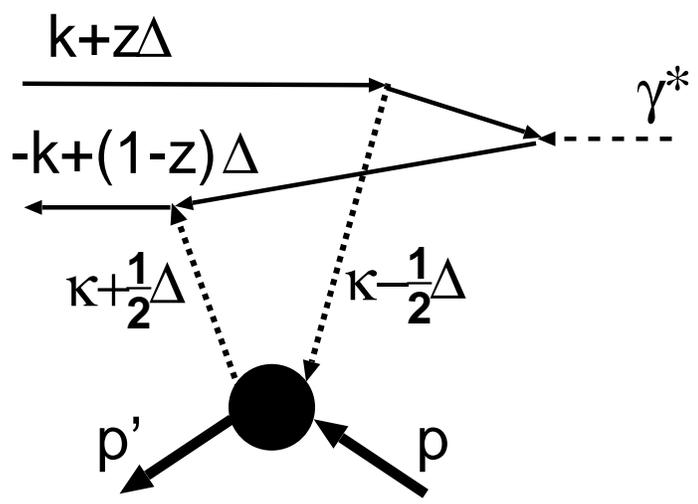
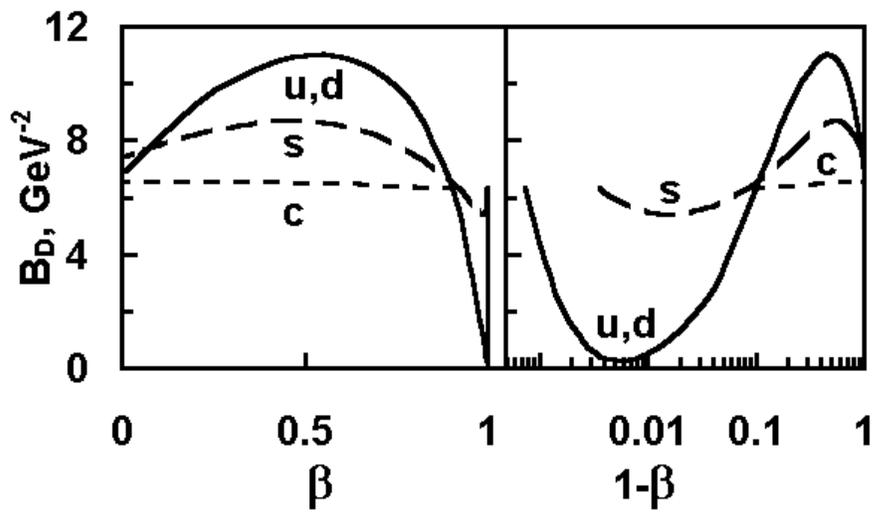
\end{document}